\documentclass[12pt]{iopart}

\usepackage{graphicx} 
\usepackage{float} 
\usepackage{amssymb}
\usepackage{url}
\usepackage{hyperref}
\newcommand{\pT} {\ensuremath{p_{\mathrm{T}}}}

\begin{document}

\title[A study of the anisotropy associated with dipole asymmetry in heavy ion collisions]{A study of the anisotropy associated with dipole asymmetry in heavy ion collisions}
\author{Jiangyong Jia$^{1,2}$, Sooraj Radhakrishnan$^1$,Soumya Mohapatra$^1$}
\address{$^1$ Department of Chemistry, Stony Brook University, Stony Brook, NY 11794, USA}
\address{$^2$ Physics Department, Brookhaven National Laboratory, Upton, NY 11796, USA}
\ead{jjia@bnl.gov,sooraj9286@gmail.com,smohapatra@gmail.com}

\begin{abstract}
The anisotropy associated with the initial dipole asymmetry in heavy ion collisions is studied via the two-particle relative azimuthal azimuthal angle ($\Delta\phi=\phi^{\mathrm a}-\phi^{\mathrm b}$) correlations, within a multi-phase transport model. For a broad selection of centrality, transverse momenta ($\pT^{\mathrm {a,b}}$) and pseudorapidity ($\eta^{\mathrm {a,b}}$), a fitting method is used to decompose the anisotropy into a rapidity-even component, characterized by the Fourier coefficient $v_1$, and a global momentum conservation component. The extracted $v_1$ values are negative for $\pT\lesssim0.7-0.9$ GeV, reach a maximum at 2-3 GeV, and decreases at higher $\pT$. The $v_1$ values vary weakly with $\eta$ and centrality, but increases with collision energy and parton cross-section. The extracted global momentum conservation component is found to depend on $\Delta\eta= \eta^{\mathrm {a}}-\eta^{\mathrm {b}}$ for $|\Delta\eta|<3$. 
\end{abstract}

\submitto{\JPG}
\maketitle 
\section{Introduction}
A large azimuthal anisotropy is observed for particle production in heavy ion collisions at the Relativistic Heavy Ion Collider (RHIC) and the Large Hadron Collider (LHC). This anisotropy has been successfully described in terms of pressure-driven collective flow, which converts the spatial asymmetries of the initially created matter into transverse momentum ($\pT$) space~\cite{Voloshin:2008dg}. Such anisotropy can be expressed as Fourier series in azimuthal angle ($\phi$):
\begin{equation}
\label{eq:vn1} \frac{dN}{d\phi} \propto 1 + 2\sum_{n=1}^{\infty}v_n(\pT)\cos n(\phi-\Phi_n),
\end{equation}
where $v_n$ and $\Phi_n$ represent the magnitude and direction of $n^{\mathrm{th}}$-order flow respectively. Asymmetries giving rise to non-zero $v_n$ are associated with either an average shape (for $n=2$) or shapes due to the spatial fluctuations of the participating nucleons~\cite{Miller:2003kd,Manly:2005zy,Alver:2010gr,Alver:2010dn,Staig:2010pn,Teaney:2010vd}; the $n=1$ term is known as dipole asymmetry. Due to the approximate boost invariance of the created matter, the $v_n$ are expected to be even functions in rapidity (rapidity-even)~\cite{Gardim:2011qn,Bozek:2010vz}. However, a small rapidity-odd $v_n$ component, especially for $n=1$, may arise from the ``sideward'' deflection of the created matter~\cite{Voloshin:2008dg}. This rapidity-odd $v_1$ is expected to be very small for the rapidity range and the collision energies considered here~\cite{Back:2005pc,Abelev:2008jga,Selyuzhenkov:2011zj}, and is not considered in this paper.

The harmonics in Eq.~\ref{eq:vn1} is also reflected in the two-particle correlation (2PC) in $\Delta\phi=\phi^{\mathrm{a}}-\phi^{\mathrm{b}}$:
\begin{eqnarray}
\label{eq:vnna}
 \frac{dN_{\mathrm{pairs}}}{d\Delta\phi} \propto 1+2\sum_{n=1}^{\infty}v_{n,n}(\pT^{\mathrm{a}},\pT^{\mathrm b}) \cos n\Delta\phi\;,
\end{eqnarray} 
with
\begin{eqnarray}
\label{eq:vnnb}
v_{n,n}(\pT^{\mathrm a},\pT^{\mathrm b}) = v_n(\pT^{\mathrm a})v_n(\pT^{\mathrm b})\;.
\end{eqnarray} 
Recent studies at the LHC show that the latter factorization is valid for $n\ge2$ at low $\pT$ for particle pairs with a large rapidity gap~\cite{Aamodt:2011by,CMS:2012wg,ATLAS:2012wg}, but fails for $n=1$ over the entire $\pT$ range. This is understood to be due to global momentum conservation for a finite multiplicity system, which may induce a negative dipole component in Eq.~\ref{eq:vnnb}~\cite{Borghini:2000cm,Borghini:2002mv,Borghini:2006yk,Luzum:2010fb,Gardim:2011qn}:
\begin{eqnarray}
\label{eq:v1a}
v_{1,1}(\pT^{\mathrm a},\pT^{\mathrm b}) = v_1(\pT^{\mathrm a})v_1(\pT^{\mathrm b})-c\pT^{\mathrm a}\pT^{\mathrm b},\;\;c= \frac{1}{KM\langle \pT^2\rangle}\label{eq:v1b}
\end{eqnarray}
where $M$ is the total multiplicity in the event and $K$ is the fraction of particles that are correlated via momentum conservation, and hence the parameter $c$ has a dimention of GeV$^{-2}$. Recent studies by the ATLAS Collaboration~\cite{ATLAS:2012wg} and Retinskaya~{\it et.al.}~\cite{Retinskaya:2012ky} shows that $v_{1,1}$ can indeed be described by Eq.~\ref{eq:v1a} over broad ranges of $\pT^{\mathrm {a,b}}$. 

Many models based on ideal/viscous hydrodynamics or transport theory have been developed to describe the $v_n$ and $v_{n,n}$ data~\cite{Gardim:2011qn,Xu:2011jm,Xu:2011fe,Xu:2010du,Ma:2010dv,Schenke:2011bn,Qiu:2011iv,Qiu:2011hf,Takahashi:2009na}. They show that sizable higher-oder flow ($n\ge2$) can be generated by including fluctuations of the initial geometry. In some cases, reasonable agreement with the $v_n$ data has been achieved. However, due to the complication associated with the global momentum conservation, most of these model attempts have avoided a prediction for $v_1$. Recently, Gardim {\it et.al}~\cite{Gardim:2011qn} performed the first calculation of the rapidity-even $v_1$ in a hydrodynamic calculations; a modified event plane (EP) method was also proposed~\cite{Luzum:2010fb}. The $v_1(\pT)$ was predicted to change sign at low $\pT$, due to the constraint that the total transverse momentum of the system arising from dipolar flow should vanish, i.e. $\int d\pT \pT v_1(\pT) = 0$~\cite{Gardim:2011qn}. This sign-flip can lead to rather complicated dependence of $v_{1,1}$ on $\pT^{\mathrm a}$ and $\pT^{\mathrm b}$ (see Eq.~\ref{eq:v1b}). This complicated pattern has been observed in the ATLAS data, and is expected to also exist in transport models. Furthermore, the systems that are related via momentum conservation could be smaller than the whole event (for example a dijet system)~\cite{Chajecki:2008yi}. Therefore, the value of $c$ (Eq.~\ref{eq:v1b}) may depend on other kinematic variables such as the relative pseudorapidity ($\Delta\eta$) of the particle pairs. Hence, it is advantageous to determine $c$ via the fit method used by the ATLAS Collaboration, instead of calculating it directly.

In this work, we present a quantitative study of the rapidity-even $v_1(\pT,\eta)$ using a generalization of the ATLAS fitting method within the AMPT (A Multi Phase Transport) model~\cite{Lin:2004en}. The AMPT model combines the initial condition from HIJING~\cite{Gyulassy:1994ew} and final state interaction via a parton and hadron transport model. The outputs are complete monte-carlo events that naturally contains various flow and non-flow effects, including the global momentum conservation. We shall demonstrate that the $v_{1,1}$ from AMPT has similar features as the experimental data, and can be well described by Eq.~\ref{eq:v1a}, lending support to the procedure used by ATLAS. We then make quantitative predictions on how $v_1$ varies with $\pT$, $\eta$, centrality, collision energy and parameters in the partonic transport, as well as to compare with the ATLAS data.

\section{Analysis method}
The initial condition from AMPT is seeded by a Glauber model for nucleon-nucleon scatterings, hence it contains the fluctuations which generate the event-by-event spatial asymmetries. The nucleon-nucleon scatterings provide both minijet partons for hard scattering and strings for soft coherent interactions; these strings are then converted into soft partons via a string melting scheme. The parton and hadron transport is responsible for transforming the initial asymmetries into the momentum anisotropy. The parton transport include only elastic scattering whose cross-sections are controlled by the values of the strong coupling constant ($\alpha_s$) and the Debye screening mass ($\mu$)~\cite{Lin:2004en}: 
\begin{eqnarray}
\label{eq:cros}
\frac{d\sigma}{dt}\approx \frac{9\pi\alpha_s^2}{2(t-\mu^2)^2},\;\;\; \sigma\approx9\pi\alpha_s^2/(2\mu^2)
\end{eqnarray} 
where $t$ is the Mandelstam variable for four momentum transfer. The total cross-section can be increased either with a larger $\alpha_s$ or a smaller value of $\mu$. However changing $\alpha_s$ is more effective for momentum dissipation than changing $\mu$, as decreasing $\mu$ includes only softer scattering (Eq.~\ref{eq:cros}). Partons are recombined into hadrons at the freezeout, followed by a hadronic transport. 

AMPT events are generated for two different $\alpha_s$ values for parton transport: set-A with $\alpha_s=0.47$ and set-B with $\alpha_s=0.33$ from~\cite{Xu:2011fe}~\footnote{set-A and set-B also uses different parameterization of the Lund fragmentations in HIJING, which affect mainly the total multiplicity but not $v_1$.}. We vary the total cross-section from 1.5 mb to 10 mb by adjusting $\mu$ according to~\cite{Lin:2004en}, separately for set-A and set-B. Varying $\alpha_s$ and $\mu$ allows us to study how $v_1$ depends on the strength of final state interactions. These studies are carried out at RHIC energy (Au+Au at $\sqrt{s_{NN}}=0.2$ TeV) and/or LHC energy (Pb+Pb at $\sqrt{s_{NN}}=2.76$ TeV) for the impact parameters $b=4$~fm and $b=8$~fm, corresponding to approximately 0-10\% and 30-40\% most central events, respectively.

We follow the procedure of extracting $v_1$ as documented in~\cite{ATLAS:2012wg}. First, the correlation function for a given $\pT^{\mathrm a}$, $\pT^{\mathrm b}$ and $\Delta\eta$ range is constructed from all final state particles, and $v_{1,1}(\pT^{\mathrm a}$, $\pT^{\mathrm b},\Delta\eta)$ values are calculated. Second, a least-square minimization of these $v_{1,1}$ values following Eq.~\ref{eq:v1a} is used to obtain $v_1^{\mathrm {Fit}}(\pT)$ for a given $\Delta\eta$ range, where the $v_1^{\mathrm {Fit}}(\pT)$ function is defined by its values at 9 $\pT$ points (0.5,0.7,0.9,1.1,1.5,2.0,3.0,3.8,4.6 GeV) via a cubic-spline interpolation procedure from ROOT~\cite{inter}.

\section{Results and discussion}
\begin{figure}[!t]
\includegraphics[width= 0.5\linewidth]{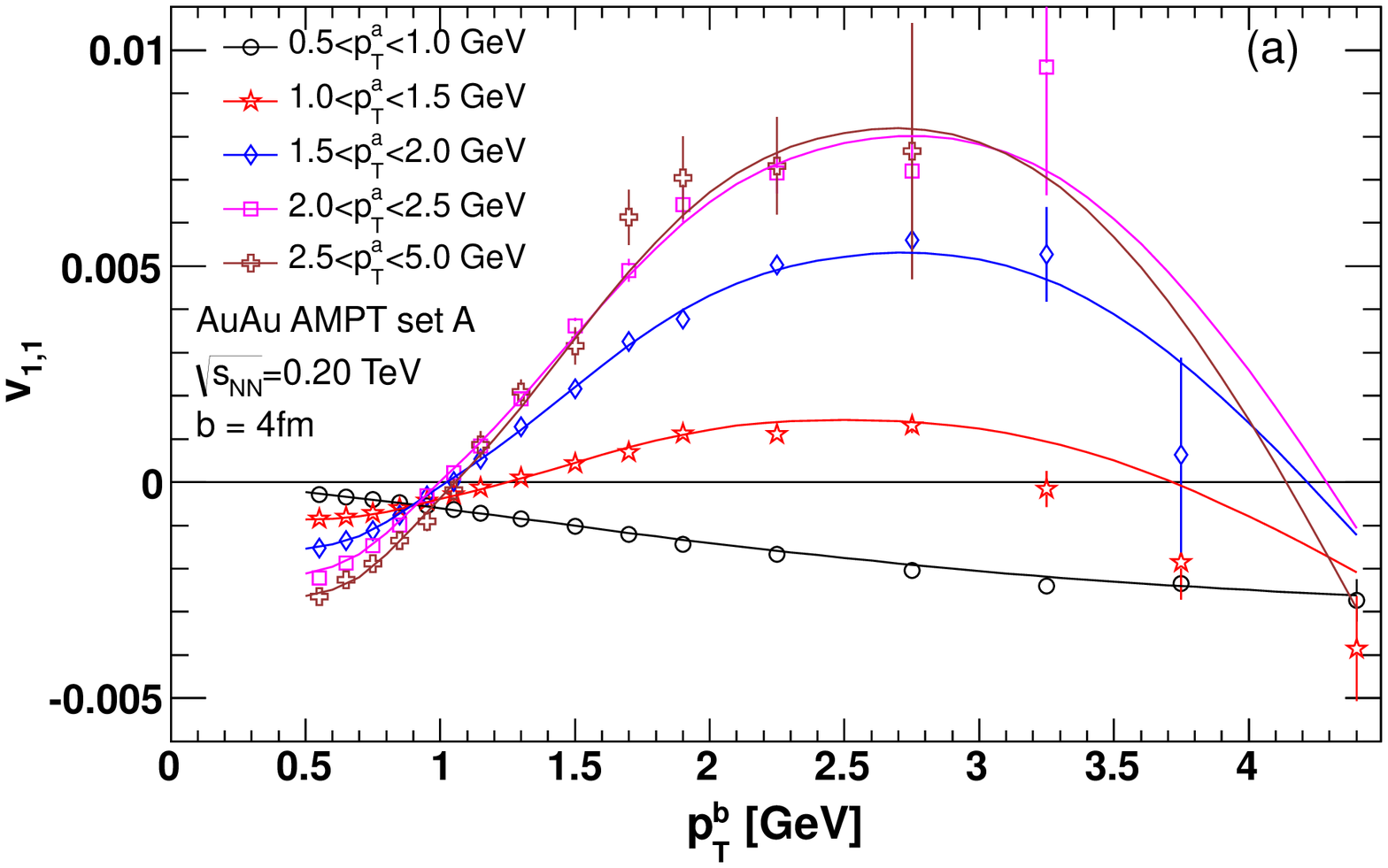}\includegraphics[width= 0.5\linewidth]{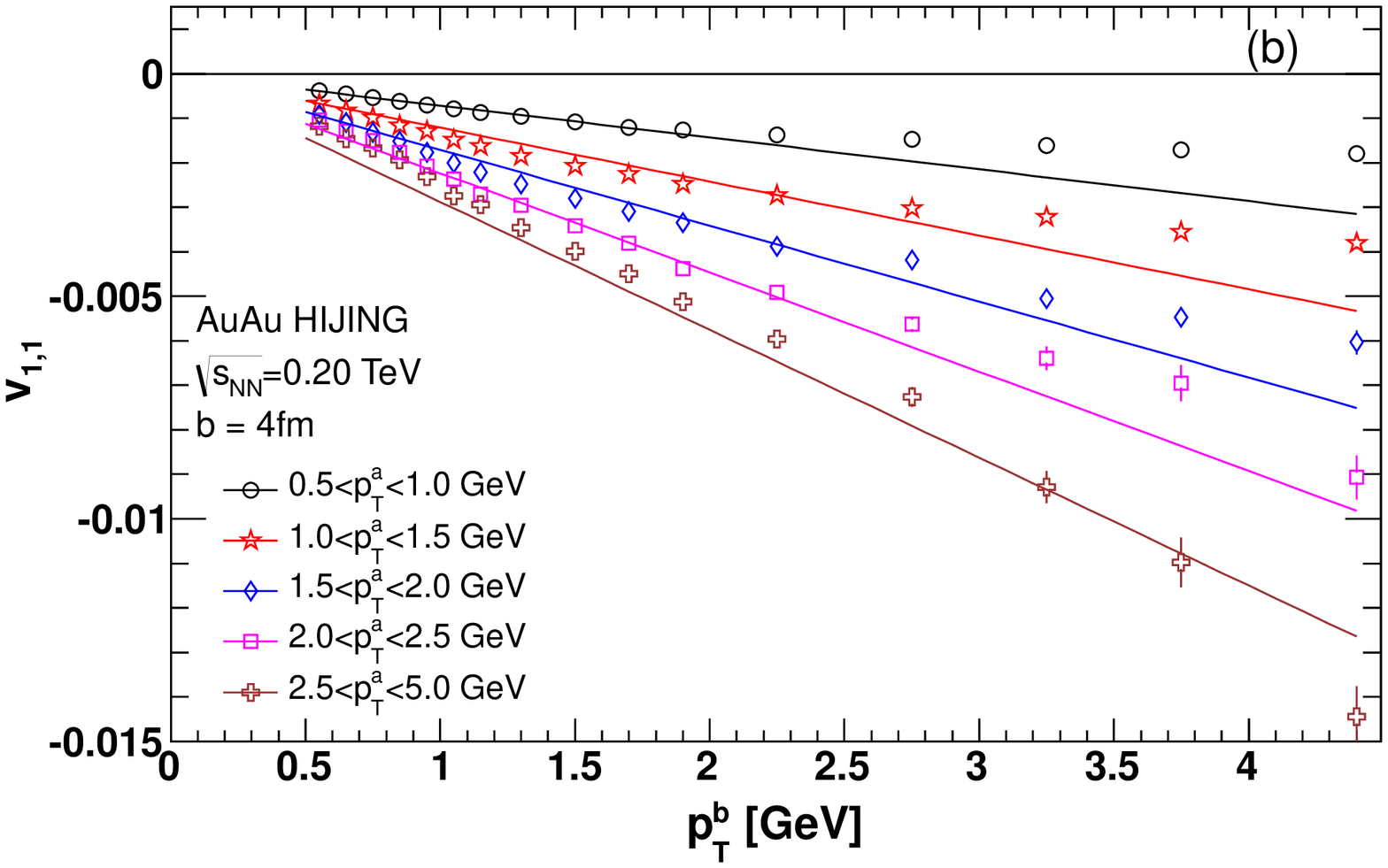}
\includegraphics[width= 0.5\linewidth]{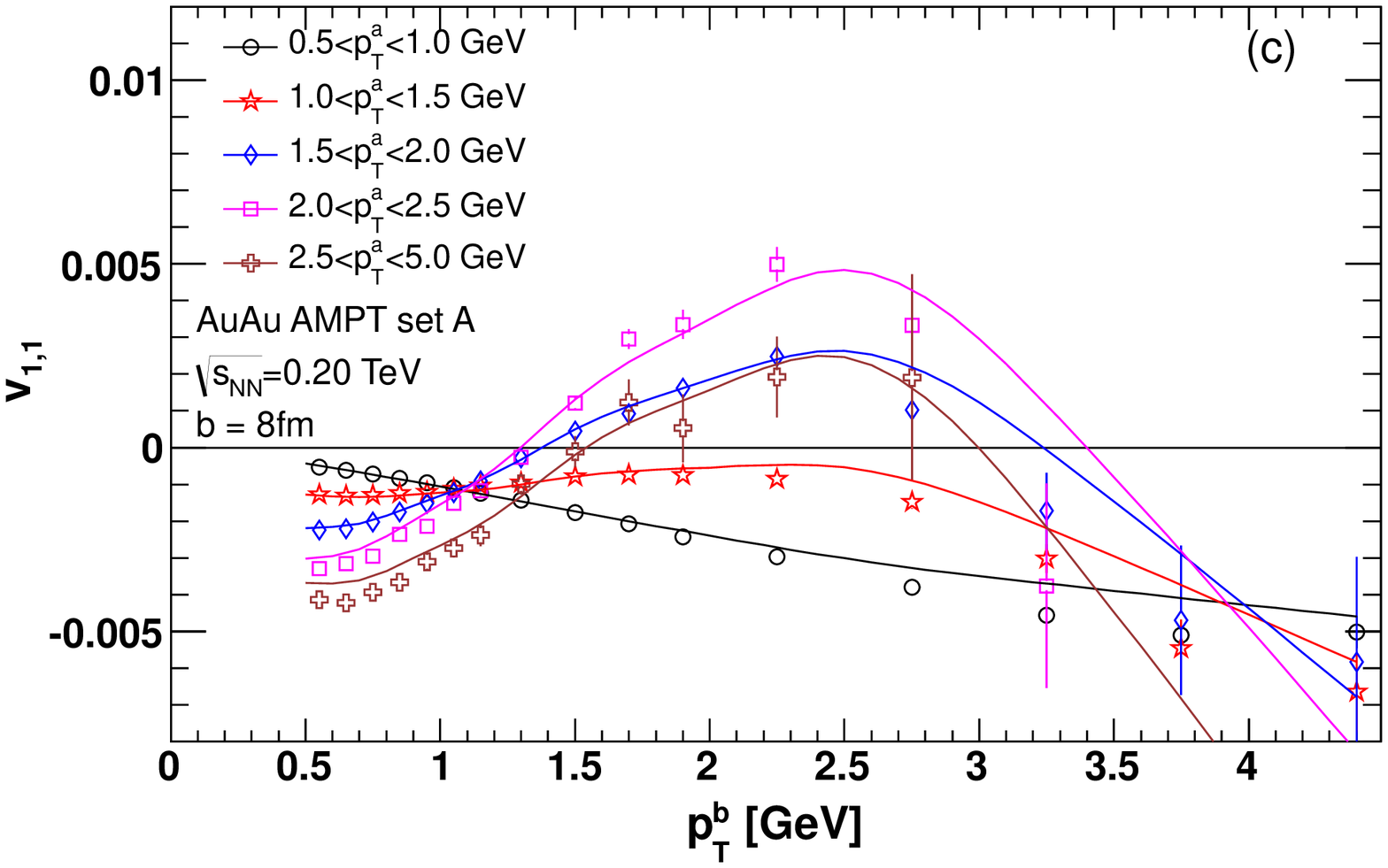}\includegraphics[width= 0.5\linewidth]{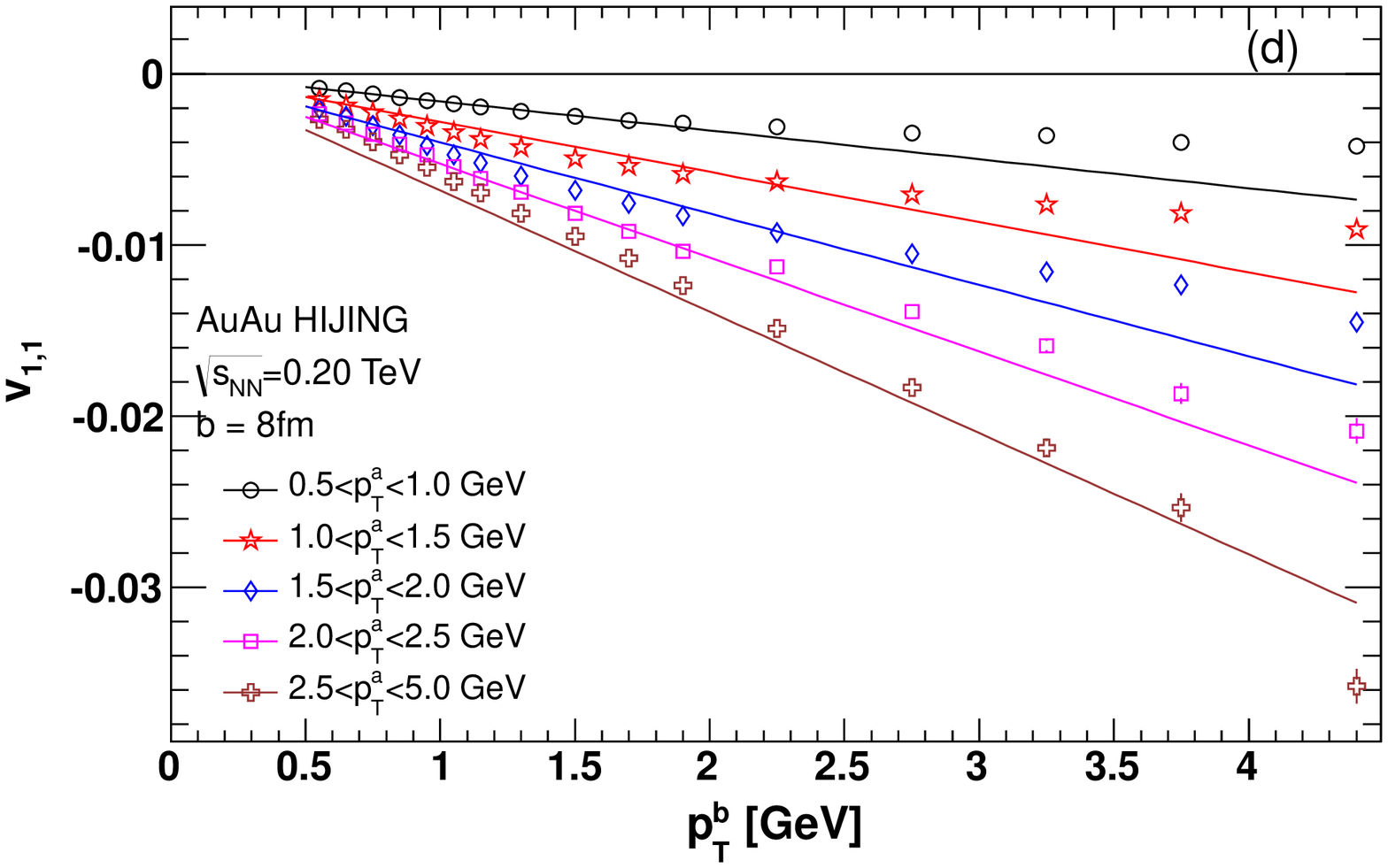}
\caption{\label{fig:1} (Color online) $v_{1,1}(\pT^{\mathrm a},\pT^{\mathrm b})$ (symbols) from AMPT (left panels) and HIJING (right panels) for $b=4$~fm (top panels) and $b=8$~fm in AuAu collisions at $\sqrt{s_{NN}}=0.2$~TeV. Lines in the left (right) panels are result of a global fit to Eq.~\ref{eq:v1a} (a fit to pure momentum conservation component: $c\pT^{\mathrm a}\pT^{\mathrm b}$).}
\end{figure}

Figure~\ref{fig:1} (a) and (c) show examples of such a study at RHIC energy, with the fit overlaid with the $v_{1,1}$ from AMPT for set-A with a 10mb partonic cross-section. A pseudorapidity gap of 1.5 is imposed to suppress the short range correlations. The $v_{1,1}$ values for different $\pT^{\mathrm a}$ selections cross each other and change sign as one increase $\pT^{\mathrm {a,b}}$, a feature qualitatively similar to the ATLAS data. 

\begin{figure}[!t]
\centering
\includegraphics[width= 0.8\linewidth]{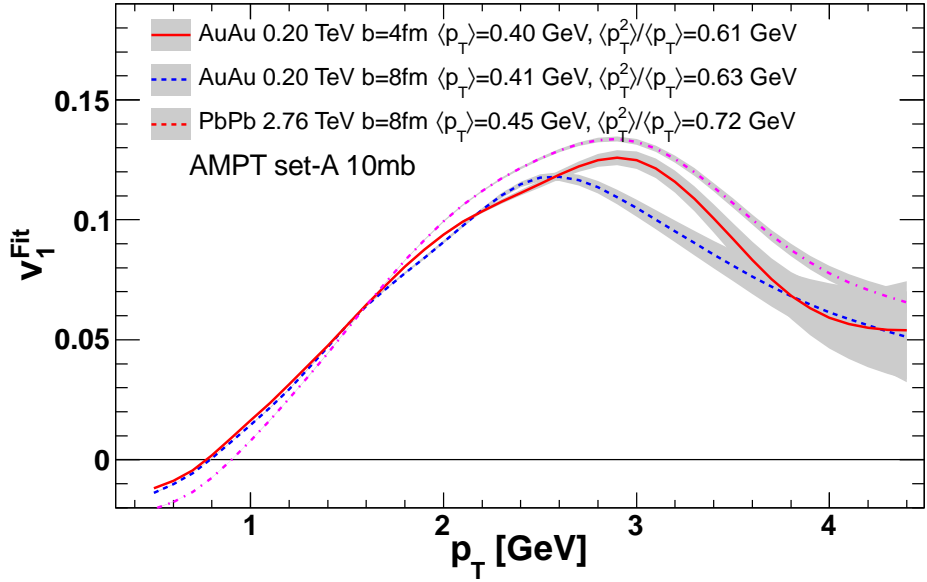}
\caption{(Color online) $v_1^{\mathrm {Fit}}$ vs $\pT$ at RHIC energy and LHC. The sign $v_1$ is chosen to be positive at high $\pT$. Shaded bands represent the systematic uncertainties evaluated using procedure similar to that of the ATLAS data analysis.}
\label{fig:2}
\end{figure}

Figure~\ref{fig:1} (b) and (d) show the $v_{1,1}$ values obtained from HIJING, which has no final state interaction. They vary nearly linearly with $\pT^{\mathrm a}\times\pT^{\mathrm b}$, consistent with behavior expected for global momentum conservation. This suggests that the complex dependence of the $v_{1,1}$ on $\pT^{\mathrm {a,b}}$ in the left panels is a natural consequence of final state interactions. Note that the $v_{1,1}$ values for $0.5<\pT^{\mathrm a}<1$ GeV are similar between AMPT and HIJING, this is because $v_{1}$ is very close to zero in this $\pT$ range (see Figure~\ref{fig:2}), hence $v_{1,1}$ values are dominated by momentum conservation term according to Eq.~\ref{eq:v1b}. 

Figure~\ref{fig:2} shows $v_1^{\mathrm {Fit}}(\pT)$ extracted from the fit. The $v_1^{\mathrm {Fit}}$ function crosses zero at low $\pT$, with the crossing point of $p_{\mathrm{T},0}=0.7-0.8$ GeV at RHIC energy and 0.9 GeV at LHC energy, respectively. Since the total transverse momentum arising from dipolar flow is expected be zero, i.e. $\int d\pT \pT v_1(\pT) = 0$~\cite{Gardim:2011qn}, the change of $p_{\mathrm{T},0}$ mainly reflects the change of the $\langle\pT\rangle$ and $\langle\pT^2\rangle$ (value shown in the legend). One can show that if $v_1(\pT)$ is a linear function of $\pT$, the crossing point is $p_{\mathrm{T},0}=\langle\pT^2\rangle/\langle\pT\rangle$. However, a change in the $\langle\pT\rangle$ and $\langle\pT^2\rangle$ should have little influence on the $v_1$ values at high $\pT$; there the $v_1$ values are more sensitive to transport properties of the medium. The $v_1^{\mathrm {Fit}}(\pT)$ increases with $\pT$ to a value $\sim0.1$ at 2-3 GeV, and then decreases at higher $\pT$. The $v_1^{\mathrm {Fit}}$ values for $b=4$~fm and $b=8$~fm are comparable, consistent with the lack of centrality dependence in the initial dipole asymmetry~\cite{Staig:2010pn,Teaney:2010vd}. The $v_1^{\mathrm {Fit}}$ value at LHC energy is larger, consistent with a stronger collective flow at LHC.

\begin{figure}[!t]
\centering
\includegraphics[width= 1\linewidth]{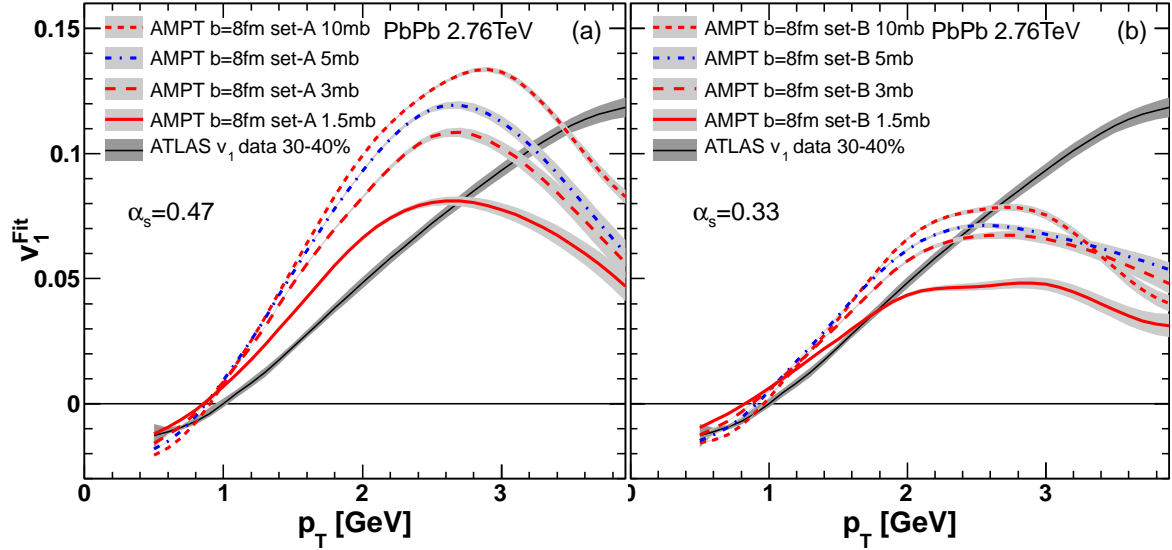}
\caption{(Color online) $v_1^{\mathrm {Fit}}$ vs $\pT$ for set-A (left panels) and set-B (right panels) for impact parameter $b=8$~fm, and compared with ATLAS Pb+Pb data. Shaded bands represent the systematic uncertainties.}
\label{fig:3}
\end{figure}

Figure~\ref{fig:3} shows $v_1^{\mathrm{Fit}}(\pT)$ for set-A and set-B at different parton cross-sections tuned by changing $\mu$. The magnitude of $v_1^{\mathrm {Fit}}(\pT)$ clearly grows with increasing cross-section. However the $v_1$ values are more sensitive to changes in $\alpha_s$: a change of $\alpha_s=0.47$ for set-A to 0.33 for set-B almost leads to a 60\% increase of the peak value of $v_1$. Results for set-B with a 1.5~mb cross-section agree better with the ATLAS data at low $\pT$. However both sets fail to describe the trend of the data for $\pT>3$ GeV. This reflects a limitation in the AMPT model for predicting the behavior of $v_n$ at high $\pT$: the jet quenching physics is important at high $\pT$, but is not modeled correctly in AMPT. Similar studies for $v_2$ also found that the set-B 1.5~mb has the best agreement with the data for $\pT<2$ GeV~\cite{Xu:2011fe}.

The original fitting method of ATLAS can be generalized to measure the $\eta$ dependence of $v_1$ in AMPT. We first perform the 2PC analysis in bins of $\pT^{\mathrm a},\pT^{\mathrm b},\eta^{\mathrm a}$, and $\eta^{\mathrm b}$. The resulting 4-D dataset $v_{1,1}(\pT^{\mathrm a},\pT^{\mathrm b},\eta^{\mathrm a},\eta^{\mathrm b})$ is then parameterized by:
\begin{eqnarray}
  v_{1,1}(\pT^{\mathrm a},\pT^{\mathrm b},\eta^{\mathrm a},\eta^{\mathrm b}) = v_1^{\mathrm {Fit}}(\pT^{\mathrm a},\eta^{\mathrm a})v_1^{\mathrm {Fit}}(\pT^{\mathrm b},\eta^{\mathrm b})-c(\eta^{\mathrm a},\eta^{\mathrm b})\pT^{\mathrm a}\pT^{\mathrm b}\;,\label{eq:v1c}
\end{eqnarray}
to directly obtain $v_1^{\mathrm {Fit}}(\pT,\eta)$. A minimum $|\Delta\eta|$ separation for the pairs is imposed to suppress the short range non-flow correlations. Therefore the total number of $v_{1,1}$ data points used in the fit depends on this gap requirement and the binning in $\pT$ and $\eta$. 

In the default fit, both $v_1^{\mathrm {Fit}}$ and $c$ functions are assumed to be symmetric in $\eta$ i.e. $v_1^{\mathrm {Fit}}(\eta)=v_1^{\mathrm {Fit}}(-\eta)$ and $c(\eta^{\mathrm a},\eta^{\mathrm b})=c(-\eta^{\mathrm a},-\eta^{\mathrm b})$. In addition, we also have $c(\eta^{\mathrm a},\eta^{\mathrm b})=c(\eta^{\mathrm b},\eta^{\mathrm a})$. We divide the $|\eta|<3$ into 12 equally spaced $\eta$ bins and use the same 9 bins in $\pT$ as for Fig~\ref{fig:1} and \ref{fig:2}. The fit uses a $12\times9=108$ mesh in $(\eta,\pT)$ space with a simple linear interpolation, which has 54 independent mesh nodes due to symmetry in $\eta$. This together with 24 parameters for $c$, gives a total of 78 fitting parameters. The quality of the fit is generally comparable to those shown in Fig.~\ref{fig:1}.

\begin{figure}
\centering
\includegraphics[width= 1\textwidth]{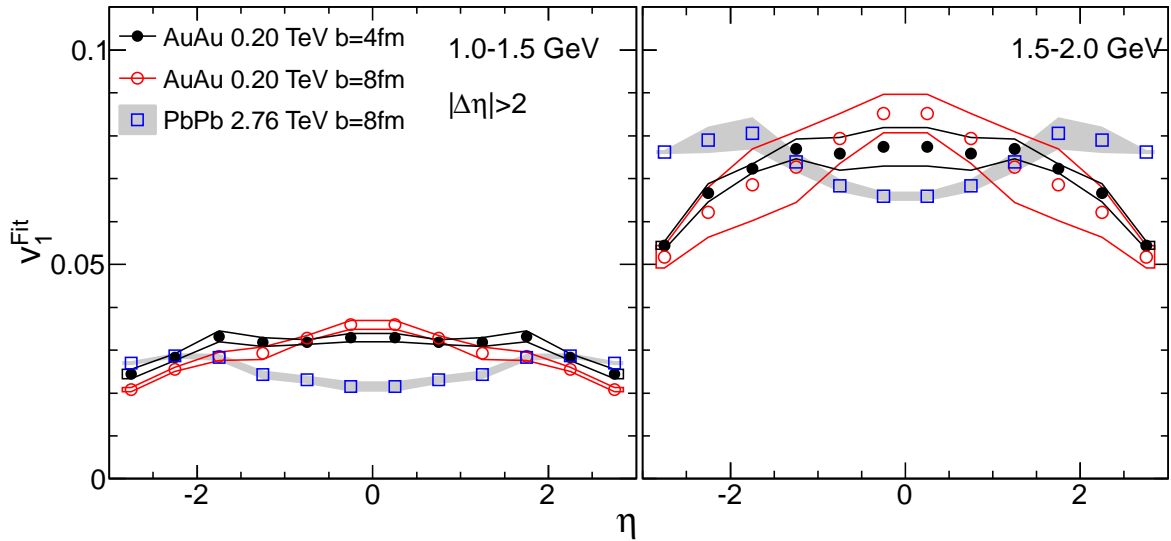}
\caption{$v_1^{\mathrm{Fit}}$ vs. $\eta$ for two $\pT$ ranges. They are determined from global fit of $v_{1,1}$ values for $|\Delta\eta|>2$. Shaded bands and contours represent the systematic uncertainties.}
\label{fig:4}
\end{figure}

\begin{figure}
\centering
\includegraphics[width= 0.5\textwidth]{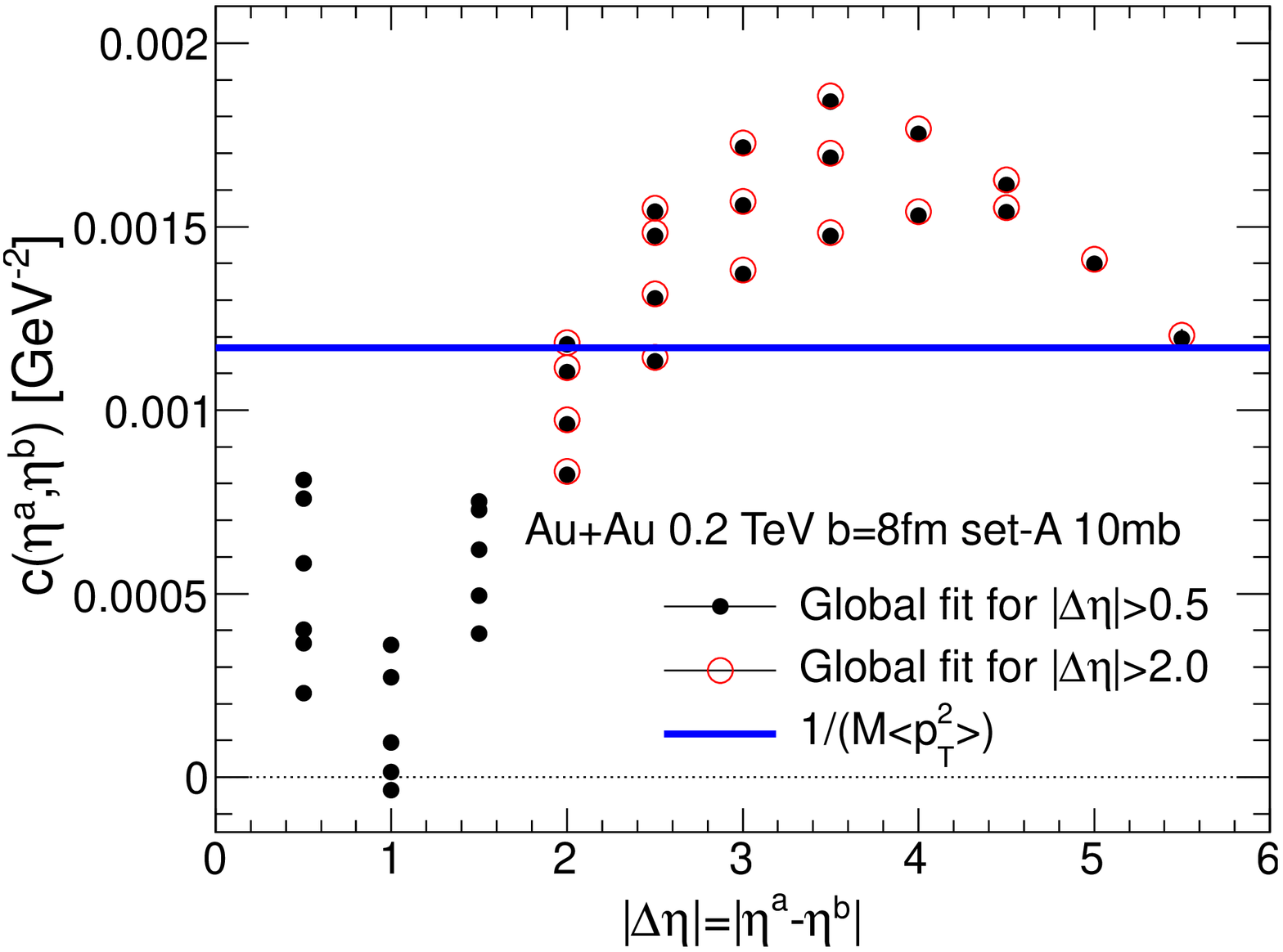}\hspace*{-0.3cm}\includegraphics[width= 0.5\textwidth]{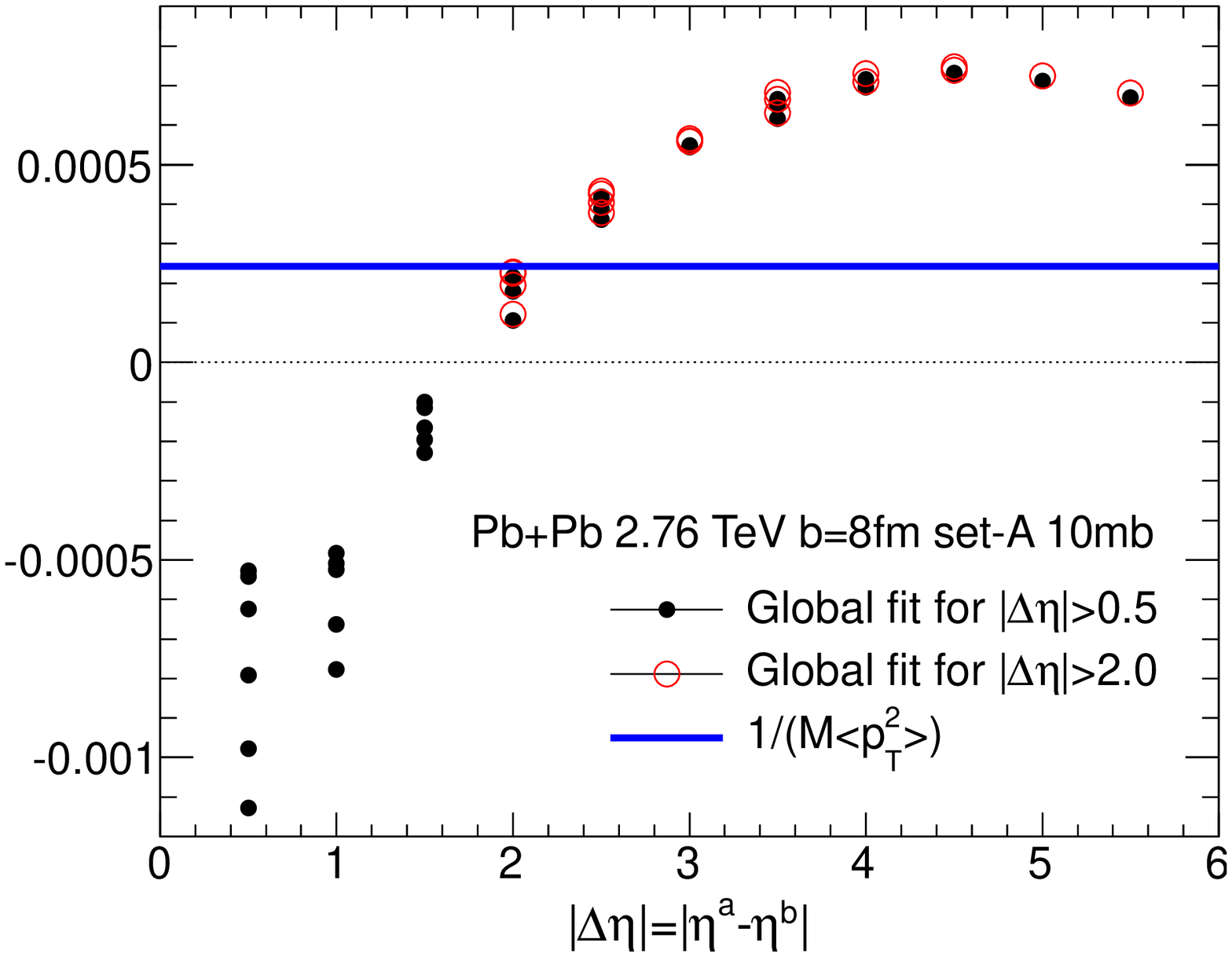}
\caption{$c(\eta^{\mathrm a},\eta^{\mathrm b})$ vs. $|\eta^{\mathrm a}-\eta^{\mathrm b}|$ for $\Delta\eta>0.5$ and $\Delta\eta>2$ obtained from the Global fit. Data points at given $|\Delta\eta|$ value corresponds to different $\eta^{\mathrm a}$ and $\eta^{\mathrm b}$ combinations.}
\label{fig:5}
\end{figure}

Figure~\ref{fig:4} shows the $\eta$ dependence of the $v_1$ obtained this way for two $\pT$ bins. The error bands account for the statistical uncertainty as well as the variation of the fit result by changing the $|\Delta\eta|$ gap from 0.5 to 2.5. The $v_1^{\mathrm {Fit}}(\eta)$ generally shows a weak $\eta$-dependence at RHIC energy, especially in the more central collisions. At LHC energy, however, the $v_1^{\mathrm {Fit}}(\eta)$ shows a small dip at mid-rapidity. The physics behind this dip is currently under investigation, but probably is related to the stronger longitudinal flow at higher collision energy. Figure~\ref{fig:5} shows the values of $c(\eta^{\mathrm a},\eta^{\mathrm b})$ plotted as a function of $|\Delta\eta|$ for $b=8$~fm at RHIC and LHC. They fall approximately on a common curve for $|\Delta\eta|>2$, suggesting that $c$ nearly depends only on $|\Delta\eta|$ for pairs with large rapidity separation. The values of $c$ at large $|\Delta\eta|$ are not sensitive to the $|\Delta\eta|$ gap used in the global fit (compare the solid and open symbols), and they are three times larger than that calculated via Eq.~\ref{eq:v1b} assuming $K=1$ at LHC (right panel), but are comparable at RHIC (left panel).

In the original functional Eq.~\ref{eq:v1a}, parameter $c$ is assumed to be a single number for each centrality class. Figure~\ref{fig:5} clearly show that this is not the case in AMPT, and the global momentum conservation contribution changes with $|\Delta\eta|$. Some of this dependence is due to the influence of the short range correlations such as jet fragmentation and resonance decay, which tend to give a negative contribution to $c$ at small $|\Delta\eta|$. However, the values of $c$ still change with $|\Delta\eta|$ up to $|\Delta\eta|=3$ then seem to flatten out at higher $|\Delta\eta|$, but at a level much bigger than that predicted by Eq.~\ref{eq:v1b}. This may suggest that evaluating $c$ with all particles in the event might be incorrect. In the modified EP method of Ref.~\cite{Luzum:2010fb}, a $\pT$ dependent weight $w = \pT-\langle\pT^2\rangle/\langle\pT\rangle$ is calculated for each particle used in the EP calculation. Note that if $c$ depends on $\eta$ of the two particles, $\langle\pT^2\rangle$ and $\langle\pT\rangle$ need to be calculated in bins of $\eta$, instead of a single number calculated over the full $\eta$ range of the EP. In other words, the weight is a function of $\pT$ and $\eta$: $w(\pT,\eta) = \pT-\langle\pT^2\rangle(\eta)/\langle\pT\rangle(\eta)$. 

\section{Summary}
In summary, we study the $v_1$ associated with the initial dipole asymmetry in the AMPT model using $v_{1,1}=\langle\cos\Delta\phi\rangle$ obtained from the two-particle correlation technique. A global fit of the $v_{1,1}$ using the procedure from the ATLAS Collaboration~\cite{ATLAS:2012wg} allows the simultaneous extraction of $v_1(\pT)$ and the global momentum conservation component. The excellent fit demonstrates that the complex pattern of the $v_{1,1}$ data from the model, which are similar to the ATLAS data, is a natural consequence of the interplay between final state interaction and global momentum conservation. The extracted $v_1$ function is negative at low $\pT$, and increases with $\pT$ until 2-3 GeV, qualitatively agreeing with the behavior predicted by the hydrodynamic models. The $v_1$ function shows very little centrality dependence but increases with collision energy and parton cross-sections, suggesting that its magnitude increases with the strength of the final state interactions. By choosing the parameters in AMPT carefully, reasonable agreement with the ATLAS data can be achieved for $\pT<2$ GeV, but not at higher $\pT$. The fitting method is extended to directly extract the $\eta$ dependence of $v_1$ and the global momentum conservation component (Eq.~\ref{eq:v1c}). The extracted $v_1$ shows a weak dependence on $\eta$. The coefficient of global momentum conservation component is found to depend on $\Delta\eta$ and differs from that given by Eq.~\ref{eq:v1b}. This extended method should allow the extraction of the $v_1(\pT,\eta)$ and validation of the EP method (It requires detector acceptance to $\pT=0$, which is usually not satisfied in actual experiment).

We acknowledge valuable discussions with Matthew Luzum and Jean-Yves Ollitrault, in particular, the relation between the event plane method and two-particle correlation method. We thank Roy Lacey for a careful proofreading of the manuscript. This research is supported by NSF under grant number PHY-1019387 and PHY-1305037.

\section*{References}
\begin{thebibliography}{10}
\bibitem{Voloshin:2008dg}
  S.~A.~Voloshin, A.~M.~Poskanzer, R.~Snellings,
  arXiv:0809.2949 [nucl-ex], and references there in.

\bibitem{Miller:2003kd}
  M.~Miller and R.~Snellings,
  nucl-ex/0312008.
\bibitem{Manly:2005zy} 
  PHOBOS Collaboration,
  Nucl.\ Phys.\ A {\bf 774}, 523 (2006); Phys.\ Rev.\ Lett.\  {\bf 98}, 242302 (2007).
\bibitem{Alver:2010gr}
  B.~Alver and G.~Roland,
  Phys.\ Rev.\  C {\bf 81}, 054905 (2010)
  [Erratum-ibid.\  C {\bf 82}, 039903 (2010)].
\bibitem{Alver:2010dn}
  B.~H.~Alver, C.~Gombeaud, M.~Luzum, J.~-Y.~Ollitrault,
  Phys.\ Rev.\  C {\bf 82}, 034913 (2010).
\bibitem{Staig:2010pn}
 P.~Staig and E.~Shuryak,
 Phys.\ Rev.\  C {\bf 84}, 034908 (2011).
\bibitem{Teaney:2010vd}
  D.~Teaney and L.~Yan,
Phys.\ Rev.\  C {\bf 83}, 064904 (2011).
\bibitem{Gardim:2011qn} 
  F.~G.~Gardim, F.~Grassi, Y.~Hama, M.~Luzum and J.~-Y.~Ollitrault,
  Phys.\ Rev.\ C {\bf 83}, 064901 (2011).
\bibitem{Bozek:2010vz} 
  P.~Bozek, W.~Broniowski and J.~Moreira,
  Phys.\ Rev.\ C {\bf 83}, 034911 (2011).

\bibitem{Back:2005pc}
 PHOBOS Collaboration,
  Phys.\ Rev.\ Lett.\  {\bf 97}, 012301 (2006).
\bibitem{Abelev:2008jga}
  STAR~Collaboration, 
  Phys.\ Rev.\ Lett.\  {\bf 101}, 252301 (2008).
\bibitem{Selyuzhenkov:2011zj}
  ALICE~Collaboration,
 J.\ Phys.\ G {\bf 38}, 124167 (2011).


\bibitem{Aamodt:2011by}
ALICE Collaboration,
 Phys.\ Lett.\ B {\bf 708}, 249 (2012).
\bibitem{CMS:2012wg} 
  CMS~Collaboration,
 Eur.\ Phys.\ J.\ C {\bf 72}, 2012 (2012). 
\bibitem{ATLAS:2012wg} 
  ATLAS~Collaboration, Phys.\ Rev.\ C {\bf 86}, 014907 (2012).
\bibitem{Borghini:2000cm}
  N.~Borghini, P.~M.~Dinh, J.~-Y.~Ollitrault,
  Phys.\ Rev.\  C {\bf 62}, 034902 (2000).
\bibitem{Borghini:2002mv}
  N.~Borghini, P.~M.~Dinh, J.~-Y.~Ollitrault, A.~M.~Poskanzer, S.~A.~Voloshin,
  Phys.\ Rev.\  C {\bf 66}, 014901 (2002).
\bibitem{Borghini:2006yk}
  N.~Borghini,
  Phys.\ Rev.\  C {\bf 75}, 021904 (2007).
\bibitem{Luzum:2010fb}
  M.~Luzum and J.~Y.~Ollitrault,
  Phys.\ Rev.\ Lett.\  {\bf 106}, 102301 (2011).
\bibitem{Retinskaya:2012ky} 
  E.~Retinskaya, M.~Luzum and J.~-Y.~Ollitrault,
  Phys.\ Rev.\ Lett.\  {\bf 108}, 252302 (2012). 

\bibitem{Xu:2011jm} 
  J.~Xu and C.~M.~Ko,
  Phys.\ Rev.\ C {\bf 84}, 044907 (2011).
\bibitem{Xu:2011fe} 
  J.~Xu and C.~M.~Ko,
  Phys.\ Rev.\ C {\bf 84}, 014903 (2011),
  Phys.\ Rev.\ C {\bf 83}, 034904 (2011).

\bibitem{Xu:2010du} 
  J.~Xu and C.~M.~Ko,
  Phys.\ Rev.\ C {\bf 83}, 021903 (2011).
\bibitem{Ma:2010dv} 
  G.~-L.~Ma and X.~-N.~Wang,
  Phys.\ Rev.\ Lett.\  {\bf 106}, 162301 (2011).
\bibitem{Schenke:2011bn}
  B.~Schenke, S.~Jeon and C.~Gale,
  Phys.\ Rev.\ C {\bf 85}, 024901 (2012).

\bibitem{Qiu:2011iv}
  Z.~Qiu and U.~W.~Heinz,
Phys.\ Rev.\  C {\bf 84}, 024911 (2011).
\bibitem{Qiu:2011hf} 
  Z.~Qiu, C.~Shen and U.~W.~Heinz,
  Phys.\ Lett.\ B {\bf 707}, 151 (2012).

\bibitem{Takahashi:2009na} 
  J.~Takahashi, B.~M.~Tavares, W.~L.~Qian, R.~Andrade, F.~Grassi, Y.~Hama, T.~Kodama and N.~Xu,
  Phys.\ Rev.\ Lett.\  {\bf 103}, 242301 (2009).
\bibitem{Chajecki:2008yi} 
  Z.~Chajecki and M.~Lisa,
  Phys.\ Rev.\ C {\bf 79}, 034908 (2009).

\bibitem{Lin:2004en} 
  Z.~-W.~Lin, C.~M.~Ko, B.~-A.~Li, B.~Zhang and S.~Pal,
  Phys.\ Rev.\ C {\bf 72}, 064901 (2005).

\bibitem{Gyulassy:1994ew}
  M.~Gyulassy, X.~-N.~Wang,
  Comput.\ Phys.\ Commun.\  {\bf 83}, 307 (1994).

\bibitem{inter}
http://root.cern.ch/root/html/ROOT\_Math\_Interpolator.html
\end{thebibliography}
\end{document}